\documentclass[prl,twocolumn,amsmath,amssymb,superscriptaddress]{revtex4}
\usepackage{graphicx}
\usepackage{color}

\begin{document}

\newcommand{\ketbra}[2]{\ensuremath{|#1\rangle\langle#2|}}
\newcommand{\ket}[1]{\ensuremath{|#1\rangle}}
\newcommand{\braket}[1]{\ensuremath{\left\langle{#1}\right\rangle}}

\newcommand{\Am}{\ensuremath{\mathbf A}}
\newcommand{\Bm}{\ensuremath{\mathbf B}}
\newcommand{\Cm}{\ensuremath{\mathbf C}}
\newcommand{\Ao}{\ensuremath{\mathrm A}}
\newcommand{\Bo}{\ensuremath{\mathrm B}}
\newcommand{\Co}{\ensuremath{\mathrm C}}

\title{Dynamical creation of bosonic Cooper-like pairs}

\date{\today}

\author{Tassilo Keilmann}
\affiliation{Max-Planck-Institut f\"{u}r Quantenoptik,
Hans-Kopfermann-Str. 1, Garching, D-85748, Germany}

\author{Juan Jos{\'e} Garc{\'\i}a-Ripoll}
\affiliation{ Facultad de Ciencias F{\'\i}sicas,
  Universidad Complutense de Madrid,
  Ciudad Universitaria s/n, Madrid, E-28040, Spain}

\begin{abstract}
  We propose a scheme to create a metastable state of paired bosonic
  atoms in an optical lattice. The most salient features of this state
  are that the wavefunction of each pair is a Bell state and that the
   pair size spans half the lattice, similar to fermionic Cooper
  pairs. This mesoscopic state can be created with a dynamical process
  that involves crossing a quantum phase transition and which is
  supported by the symmetries of the physical system. We characterize
  the final state by means of a measurable two-particle correlator
  that detects both the presence of the pairs and their size.
\end{abstract}

\maketitle

Pairing is a central concept in many-body physics. It consists on
the existence of quantum or classical correlations between pairs
of components of a many-body system. The most relevant example of
pairing is BCS superconductivity. In a superconductor, an
attractive interaction causes electrons to organize into Cooper
pairs, bosonic quantum objects where electrons are perfectly
anticorrelated in momentum and in spin. In the language of second
quantization this is described by the BCS variational wavefunction
\begin{equation}
  \label{BCS}
  |\psi_{\mathrm{BCS}}\rangle =
  \prod_k  (u_k + v_k A_{k}^\dagger) \ket{0},
\end{equation}
where $A_k^\dagger \equiv c_{k\uparrow}^\dagger
c_{-k\downarrow}^\dagger$ is an operator that creates one such Cooper
pair. Remarkably, the fact that pairing occurs in momentum space means
that the constituents of the pairs are delocalized and share some
long-range correlation.

Nowadays, pairing is also a key research topic in the field of
ultracold atoms. While atomic interactions are short range, they can
be enhanced using Feschbach resonances. These resonances allow both to
create Cooper pairs of fermionic atoms \cite{regal04, zwierlein04,
  bourdel04} and to observe the crossover from these large,
delocalized objects to a condensate of bound molecular
states. Realizing similar experiments with bosons is difficult,
because attractive interactions can induce collapse. In the paper by
Paredes \cite{paredes03} it is suggested to load an optical lattice
with hard-core bosonic atoms in two internal states that attract each
other so as to induce BCS-like pairing. A different approach, followed
in Ref.~\cite{winkler06}, is to load an optical lattice with pairs of
atoms in the regime of strong repulsive interactions.  Such metastable
on-site pairs are robust and survive about $700$ ms in the lattice.

In this Letter we propose a method to create long-range paired states
of bosonic atoms using entangled states as a resource. The method uses
an optical lattice of arbitrary geometry which is loaded with
entangled bosons in an insulator state. We will consider both on-site
pairs
\begin{equation}
  \ket{\psi} \sim \prod_{i=1}^{{L}} A_{ii}^\dagger \ket{0},\,\,\,
  A_{ij} =
  \left\{
    \begin{array}{l}
      c_{i\uparrow}c_{j\uparrow} \pm c_{i\downarrow}c_{j\downarrow}\\
      c_{i\uparrow}c_{j\downarrow} + c_{j\uparrow}c_{i\downarrow}
    \end{array}
  \right.,                      %
  \label{states1}
\end{equation}
such as the ones demonstrated in Ref.~\cite{widera04} and a larger
family of singlet and triplet states
\begin{equation}
  \ket{\psi} \sim \prod_{i=1}^{L/2} A_{2i-1,2i}^\dagger \ket{0},\,\,\,
  A_{ij} =
  \left\{
    \begin{array}{l}
      c_{i\uparrow}c_{j\uparrow} \pm c_{i\downarrow}c_{j\downarrow}\\
      c_{i\uparrow}c_{j\downarrow} \pm c_{j\uparrow}c_{i\downarrow}
    \end{array}
  \right.,
  \label{states2}
\end{equation}
some of which have been created using optical superlattices
\cite{foelling07, portopairs}. We propose to dynamically increase the
mobility of the atoms, entering the superfluid regime. During this
process, the pairs will grow in size until they form a stable gas of
long-range Cooper-like pairs that span about half the lattice
size. Contrary to works on the creation of squeezed states
\cite{rodriguez06}, the evolution considered here is not adiabatic and
the survival of entanglement is ensured by a symmetry of the
interactions.

This paper is organized as follows. First, we present the Hamiltonian
for bosonic atoms which are trapped in a deep optical lattice, have
two degenerate internal states and spin independent interactions. Next,
we prove that by lowering the optical lattice and moving into the
superfluid regime, the Mott-Bell entangled states
(\ref{states1})-(\ref{states2}) evolve into a superfluid of pairs. We
then introduce two correlators that detect the singlet and triplet
pairs and their approximate size. These correlators are used to
interpret quasi-exact numerical simulations of the evolution of two
paired states as they enter the superfluid regime. Finally, we suggest
two procedures to measure these correlators and elaborate on other
experimental considerations.

We will study an optical lattice that contains bosonic atoms in two
different hyperfine states ($\sigma=\uparrow,\downarrow$). In the
limit of strong confinement, the dynamics of the atoms is described by
a Bose-Hubbard model \cite{jaksch98}
\begin{equation}
  \label{BH}
  H = -\sum_{\langle i,j\rangle,\sigma} J_{\sigma}c_{i\sigma}^\dagger c_{j\sigma}
  + \sum_{{i}\sigma\sigma'} \frac{1}{2} U_{\sigma\sigma'}
  c_{i\sigma}^\dagger c_{i\sigma'}^\dagger c_{i\sigma'} c_{i\sigma}.
\end{equation}
Atoms move on a $d$-dimensional lattice ($d=1,2,3$) jumping between
neighboring sites with tunneling amplitude $J_\sigma,$ and
interacting on-site with strength $U_{\sigma\sigma'}.$ The
Bose-Hubbard model has two limiting regimes. If the interactions are
weak, $U\ll J$, atoms can move freely through the lattice and form a
superfluid. If interactions are strong and repulsive, $U \gg J$, the
ground state is a Mott insulator with particles pinned to different
lattice sites.

As mentioned in the introduction, we want to design a protocol that
begins with a localized entangled state
(\ref{states1})-(\ref{states2}), and, by increasing the mobility of
atoms, enlarge these pairs until the final state describes a gas of
generalized Cooper pairs of bosons. In our proposal we restrict
ourselves to symmetric interaction and hopping amplitudes
\begin{equation}
  U \equiv U_{\uparrow\uparrow} = U_{\downarrow\downarrow}=
  U_{\uparrow\downarrow} \geq 0;\; J \equiv J_\uparrow=J_\downarrow \geq 0.
  \label{choice}
\end{equation}
This symmetry makes the system robust so that, even though bosons do
not stay in their ground state, they remain a coherent aggregate of
pairs, unaffected by collisional dephasing. We will formulate this
more precisely.

\textit{ Let us take an initial state of the form given by either
  Eq.~(\ref{states1}) or (\ref{states2}). If we evolve this state
  under the Hamiltonian (\ref{BH}), with time-dependent but symmetric
  interaction $U_{\sigma\sigma'}=U(t)$ and hopping $J_\sigma=J(t)$,
  the resulting state will have a paired structure at all times
  \begin{eqnarray}
    \ket{\psi(t)} = \sum_{\mathbf{k}} c(t;\mathbf{k})
    A^{\dagger}_{k_1 k_2} \ldots A^{\dagger}_{k_{2L-1} k_{2L}}
    \ket{0},
    \label{superposition}
  \end{eqnarray}
  where $c(t,\mathbf{k})$ are complex coefficients to be determined.
}

The proof of this result begins with the introduction of a set of
operators $C_{ij} := \sum_\sigma c_{i\sigma}^\dagger c_{j\sigma}$
which form a simple Lie algebra $\left[\Co_{ij},\Co_{kl}\right] =
\Co_{il}\delta_{jk} - \Co_{kj}\delta_{il}.$ The evolution
preserves the commutation relations and maps the group onto
itself. This is evident if we rewrite the Hamiltonian
\begin{equation}
  H = - J \sum_{\langle i,j\rangle} C_{ij} + \frac{U}{2}
  \sum_i (C_{ii})^2 .
   \label{BHC}
\end{equation}
The evolution operator satisfies a Schr\"odinger equation
$i\hbar\frac{d}{dt}V(t) = H(t) V(t),$ with initial condition
$V(0)=\mathbb{I}.$ Since the Hamiltonian only contains $C_{ij}$
operators we conclude that $V(t)$ is an analytic function of these
generators. Let us now focus on the evolution of state
(\ref{states2}), given by $\ket{\psi(t)} = V(t) \prod_{i=1}^{L/2}
A_{2i-1,2i}^{\dagger} \ket{0}.$ We will use the commutation relations
between the generators of the evolution and the pair operators
$[A_{ij},C_{kl}] = \delta_{ik}A_{lj} + \delta_{jk}A_{il},$ which are
valid for any of the pairs in Eq.~(\ref{states2}). Formally, it is
possible to expand the unitary operator $V(t)$ in terms of the
correlators $C_{ij}$ and commute all these operators to the right of
the A's, where we use $C_{ij}\ket{0}=0.$ After doing this one is left
with Eq.~(\ref{superposition}). A similar proof applies for the
on-site pairs (\ref{states1}) and for any initial state which already
has a paired structure (\ref{superposition}).

The previous result includes a very simple case, which is the
abrupt jump into the non-interacting regime, $U=0$. This problem
is integrable and for the initial conditions (\ref{states1}) and
(\ref{states2}) the evolved state can be written as
\begin{equation}
  \label{pairing}
  \ket{\psi(t)} = \prod_{n=1}^N
  \sum_{i,j} w(i-n,j-n,t)
  A_{ij}^\dagger   \ket{0}.
\end{equation}
The wavepackets $w(i,j,t)$ form an orthogonal set of states, initially
localized $w(i,j,0) \propto \delta_{ij}$ or $w(i,j,0) \propto
\delta_{ij+1}$ and for large times close to a Bessel function
\footnote{The normalization of $w(i,j,t)$ depends on
  the choice of the operator $A_{ij}$.}.  We remark that the
pair wavefunctions (\ref{superposition}) and (\ref{pairing}) can include
valence bond states, however represent a
generalization both in the fact that pairs can overlap or be triplets.

In a general case, computing the many-body pair wavefunction,
$c(t;\mathbf{k})$, is an open problem. Nevertheless we can prove that
the final state \textit{does not remain in the ground state in the superfluid
  regime}, no matter how slowly one changes the hopping and
interaction. For the states in (\ref{states2}) this is evident from
the lack of translational invariance. Let us thus focus on the state
(\ref{states1}) generated by $A_{ii}=c_{i\uparrow}c_{i\downarrow}$,
which has an equal number of spin-up and down particles
$N_{\uparrow,\downarrow}=N/2$. The ground state of the same sector in
the superfluid regime, $U=0$, is a number squeezed, two-component
condensate \cite{rodriguez06}
\begin{equation}
  |\psi_{\text{NN}}\rangle \propto \tilde{c}_{0\uparrow}^{\dagger N/2}
  \tilde{c}_{0\downarrow}^{\dagger N/2} \ket{0},
\end{equation}
with $\tilde c_{0\sigma} = \tfrac{1}{\sqrt{L}}\sum_{i=1}^L
c_{i\sigma}$.  Note that we can also write this ground state as an
integral over condensates with atoms polarized along different
directions
\begin{equation}
  |\psi_{\text{NN}}\rangle  \propto
  \int d\theta\, e^{-i N \theta/2}
  (\tilde{c}_{0\uparrow}^\dagger+
  e^{i\theta}\tilde{c}_{0\downarrow}^\dagger)^N\ket{0},
\end{equation}
When this state is evolved backwards in time, into the $J=0$ regime, each
condensate transforms into an insulator with different polarization
yielding
\begin{equation}
  \ket{\psi_{\text{NN}}}\stackrel{\mathrm{MI}}{\longrightarrow}
  \sum_{\vec{n}, \sum n_k=N/2} \prod_k (c_{k\uparrow}^\dagger)^{n_k}
  (c_{k\downarrow}^\dagger)^{2-n_k} \ket{0}.
\end{equation}
Since this state is not generated by the
$A_{ii}=c_{i\uparrow}c_{i\downarrow}$ operators, we conclude that this
particular state (\ref{states1}), when evolved into the superfluid,
leaves the ground state. Furthermore, since different pairs in
Eq.~(\ref{states1}) are related by global rotations, this
statement applies to all of them.

For the rest of this Letter we focus on two important states: the
triplet pairs generated on the same site \cite{widera04} and the
singlet pairs generated on neighboring sites \cite{foelling07, portopairs},
\begin{eqnarray}
  \ket{\psi_{T}} &=& \prod_{i=1}^{L} \frac{1}{2} (c_{i\uparrow}^{\dagger 2}+
  c_{i\downarrow}^{\dagger 2})\ket{0} \label{triplet},\;\mathrm{and}\\
  \ket{\psi_{S}} &=& \prod_{i=1}^{{L/2}}\frac{1}{\sqrt{2}}
  (c_{2i-1\uparrow}c_{2i\downarrow}- c_{2i-1\downarrow}c_{2i\uparrow})
  \ket{0},\label{singlet}
\end{eqnarray}
respectively. Our goal is to study the evolution of these states as
the mobility of the atoms is increased, suggesting experimental
methods to detect and characterize the pair structure.  The main tools
in our analysis are the following two-particle connected correlators
\begin{eqnarray}
  G^{T}_{ij} :=  \braket{c_{i\uparrow}^\dagger c_{j\uparrow}^\dagger
    c_{i\downarrow} c_{j\downarrow} } -
  \braket{c_{i\uparrow}^\dagger c_{i\downarrow}} \braket{c_{j\uparrow}^\dagger
    c_{j\downarrow}},  \label{correlator}\\
  G^{S}_{ij} :=  \braket{c_{i\uparrow}^\dagger c_{j\downarrow}^\dagger
    c_{j\uparrow} c_{i\downarrow} } - \braket{c_{i\uparrow}^\dagger c_{i\downarrow} }\braket{ c_{j\downarrow}^\dagger c_{j\uparrow}},\nonumber
\end{eqnarray}
combined in two different averages
\begin{equation}
  G_{\Delta\geq 0} = \frac{1}{L-\Delta}\sum_{i=1}^{{L-\Delta}} G_{i,i+\Delta},\quad
  \bar G =  \sum_{\Delta=0}^{L-1} G_{\Delta}
  \label{averages}
\end{equation}
and what we call the pair size
\begin{equation}
  R \equiv \frac{\sum_\Delta |\Delta|\times |G_\Delta|}{ \sum_\Delta |G_\Delta|}.
  \label{pair-size}
\end{equation}

A variant of the correlator $G^T$ has been used as a pairing witness
of fermions \cite{kraus07}. We expect these correlators to give
information about the pair size and distribution also in the
superfluid regime. This can be justified rigorously in the case of an
abrupt jump into the superfluid, where the pair wavepackets remain
orthogonal and where $G_\Delta$ and $R$ characterize the spread of the
wavefunctions $w(i,j,t)$. First, note that the single-particle
expectation values such as $\langle{c_{i\downarrow}^\dagger
  c_{i\uparrow}}\rangle$ are exactly zero since $N_\uparrow$ and %
$N_\downarrow$ are even for the triplet state $\psi_T$ and balanced
for the singlet state $\psi_S.$ Second, the two-particle correlators
only have nonzero contributions where the destruction and creation
operators cancelled and subsequently created the same pair. Combining
Eqs.~(\ref{correlator}) and (\ref{pairing}) gives
\begin{eqnarray}
  G^{T}_{ij} &=& \sum_n |w(i-n,j-n,t)|^2 4,\label{weights}\\
  G^{S}_{ij} &=& -\sum_n |w(j-n,i-n,t)|^2,\nonumber
\end{eqnarray}
where we have used the symmetry of the wavefunction,
$w(i,j,t)=w(j,i,t)$. Particularized to the initial states, the triplet
$\psi_T$ gives $G^{T}_{ij}=\delta_{ij},$ $G^{T}_\Delta =
\delta_{\Delta 0},$ $\bar{G}^T=1,$ and $R^T=0$, as expected from
on-site pairs. The singlet pairs described by $\psi_S$, on the other
hand, yield $G^{S}_{ij}=-\frac{1}{2}(\delta_{ij+1}+\delta_{i+1j})$,
$G^S_\Delta=-\frac{1}{2}\delta_{\Delta 1},$ $\bar{G}^S =-\frac{1}{2},$
and initially span two sites $R^S=1.$

For a realistic study of the evolved paired states we have simulated
the evolution of $\psi_T$ and $\psi_S$ under the Bose-Hubbard model as
the hopping increases nonadiabatically in time
\begin{equation}
  J(t) =  v \times (t U/\hbar) \times U,
\end{equation}
with ramp speeds $v=0.5, 1$ and 2 in adimensional units. The
simulations were performed using Matrix Product States (MPS) on
one-dimensional lattices with up to $20$ sites and open boundary
conditions \cite{ripoll07,verstraete04}. Given the small size of these
lattices, we expect these simulations to appropriately describe even
the superfluid regime, where the size of the energy gaps and the
high occupation number per site make the MPS simulation more
difficult.

\begin{figure}
  \centering
  \includegraphics[width=0.95\linewidth]{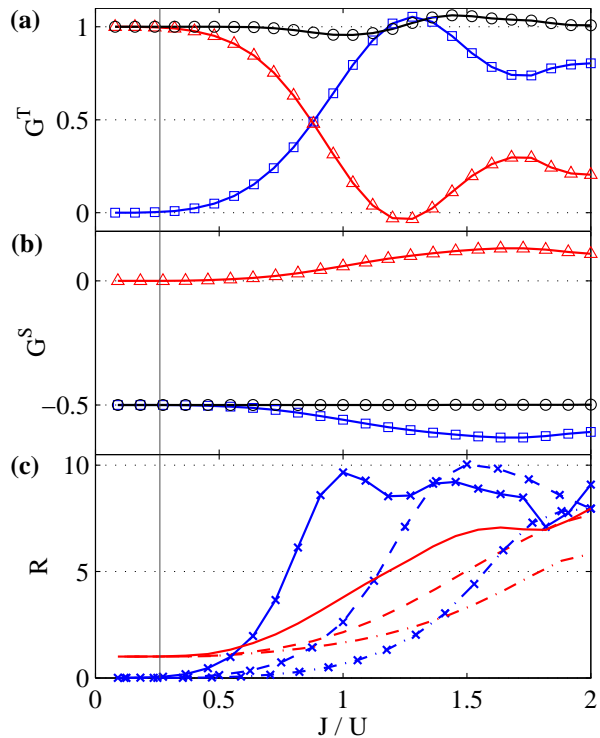}
  \caption{We plot the (a) triplet and (b) singlet correlators for the
    evolution of $\psi_T$ and $\psi_S$ respectively, in a lattice of
    $L=20$ sites. The circles, triangles and squares denote $\bar G,$
    $G_0$ and their difference. (c) Pair size $R$ for the singlet (line)
    and triplet (cross) states, for a ramp speed $v=0.5, 1$ and $2$
    (solid, dash, dash-dot). The vertical line $J/U=1/3.84$ marks the
    location of the phase transition. (Color online)}
  \label{fig:correlator}
\end{figure}

For each of these simulations we have computed all correlators and
pair sizes, values which are plotted in Fig.~\ref{fig:correlator}. Let
us begin with the triplet pairs: initially the only relevant
contribution is the short-range pair correlation $G^T_0$; then the
pair size increases monotonously from $R=0$ up to $R\sim L/2$, where
it saturates.  At this point, the pairs have become as large as the
lattice permits, given that we have uniform density and open boundary
conditions.  The singlets have a slightly different dynamics. The
antisymmetry of the spin wavefunction prevents two bosons of one pair
to share the same site and thus $R=1$ initially. However, this
antisymmetry seems also to affect the overlap between pairs, as it is
evidenced both in the slower growth $R(t)$ and in the smallness of
$G^S_0$.

Concerning the speed of the process, we have simulated ramps over a
timescale which is comparable or even shorter than the typical
interaction time, $1/U$, so that the process is definitely not
adiabatic. Nevertheless, the pairs seem to have enough time to spread
over these small lattices. Note also that the spreading of atoms
begins right after the value $J/U\simeq 1/3.84$ where the
one-dimensional Insulator-Superfluid phase transition takes place
\cite{rapsch99}.

The system of delocalized Cooper-like pairs can also be regarded as a
mean of distributing entanglement in the optical lattice.  Following
this line of thought we have used the von Neumann entropy to measure
the entanglement between two halves of the optical lattice. Apart from
a logarithmic contribution induced by the delocalization of particles,
which is also present in a single-component condensate, there is an
additional contribution caused by the spreading of pairs across the
lattice. However, this contribution does not reach a magnitude
$O(N/2)$ which corresponds to perfectly splitting $N$ pairs between
both lattice halves. We conjecture this is due to the pairs being
composed of distinguishable particles.

The pairing correlators $G^{T,S}$ can be decomposed into
density-density correlations and measured via the noise interferometry
technique developed in Ref.~\cite{altman04} and applied in
Ref.~\cite{foelling05}. To prove this, let us introduce the Schwinger
representation of angular momenta
\begin{eqnarray}
  S_{x}(i) &:=& \tfrac{1}{\sqrt{2}}(c_{i\uparrow}^\dagger c_{i\downarrow} +
  c_{i\downarrow}^\dagger c_{i\uparrow})\label{schwinger}\\
  S_{y}(i) &:=& \tfrac{1}{\sqrt{2}}(c_{i\uparrow}^\dagger c_{i\downarrow} -
  c_{i\downarrow}^\dagger c_{i\uparrow})\nonumber\\
  S_z(i) &:=& \tfrac{1}{2}(c_{i\uparrow}^\dagger c_{i\uparrow} -
  c_{i\downarrow}^\dagger c_{i\downarrow})=
  \tfrac{1}{2}(n_{i\uparrow}- n_{i\downarrow}).\nonumber
\end{eqnarray}
We will focus on the real part of the correlation matrices, defined as
$\tilde G_{ij} = 2\mathrm{Re}(G_{ij}).$ For the family of states
considered here this matrix is the same as the original one [See
Eq.~(\ref{weights})] and it is related to simple spin correlations
\begin{eqnarray}
\tilde G^{T,S}_{ij}  &=& \tfrac{1}{2} \braket{S_x(i)S_x(j) } \mp  \tfrac{1}{2} \braket{S_y(i)S_y(j) },\\
&-& \tfrac{1}{2} \braket{S_x(i)}\braket{S_x(j) } \pm
\tfrac{1}{2} \braket{S_y(i)}\braket{S_y(j) }.
\nonumber
\end{eqnarray}
We now introduce two global rotations in the hyperfine space of the
atoms $U_{x,y} = \exp\left[\pm i \frac{\pi}{2} \sum_k
  S_{y,x}(k)\right].$ The operators $U_x$ and $U_y$ take the $S_x$ and
$S_y$ operators into the $S_z$, respectively. These rotations can be
implemented experimentally without individual addressing. Using these
unitaries we can relate the previous spin correlations to simple
density operators. For instance
\begin{equation}
  \braket{S_x(i)S_x(j)} = \tfrac{1}{4}  \braket{U_x^\dagger
  (n_{i\uparrow}-n_{i\downarrow}) (n_{j\uparrow}-n_{j\downarrow})  U_x},
\end{equation}
shows that the $S_x S_x$ arises from all possible density
correlations after applying a $\pi/2$ pulse on the atoms.

Another possibility is to apply the ideas put forward in
Ref.~\cite{eckert07}. These methods rely on the interaction between
coherent light and the trapped atoms to map quantum fluctuations of
the atomic spin onto the light that crosses the lattice. Using this
technique it should be possible to measure both the single-particle
and the two-particle expectation values that constitute $G^{T,S}.$

Experimental imperfections are expected not to affect the nature
of the final state. The influence of stray magnetic and electric
fields can be cancelled by working with the singlet pairs, which
are insensitive to global rotations of the internal states and
have large coherence times. More important could be the influence
of any asymmetry in the interaction constants. However, assuming
this asymmetry to be of the order of 1$\%$, the effect can only be
noticeable after a time $t=100 \hbar/U$, which is longer than the
evolution times suggested here.

Summing up, in this Letter we have proposed a method to create a
generalization of long-range Cooper pairs using bosonic atoms in an
optical lattice.  We suggest to prepare a Mott insulator of Bell pairs
and dynamically melt this state into the superfluid regime. A spin
independent interaction guarantees that pairing is preserved and that
the system becomes a gas of long-range correlated pairs, which spread
over approximately half the lattice size.  This mechanism works even
if the process is not adiabatic or the initial state does not have
translational invariance. Most important, our proposal represents a
natural extension of current experiments with optical superlattices
\cite{foelling07, portopairs}.

We thank Ignacio Cirac for fruitful discussions. T.K. acknowledges
support from the Deutsche Telekom Stiftung.
J.J.G.R acknowledges financial support from the Ramon y Cajal
Program of the Spanish M.E.C. and from the Spanish projects
FIS2006-04885 and CAM-UCM/910758.


\begin{thebibliography}{17}
\expandafter\ifx\csname natexlab\endcsname\relax\def\natexlab#1{#1}\fi
\expandafter\ifx\csname bibnamefont\endcsname\relax
  \def\bibnamefont#1{#1}\fi
\expandafter\ifx\csname bibfnamefont\endcsname\relax
  \def\bibfnamefont#1{#1}\fi
\expandafter\ifx\csname citenamefont\endcsname\relax
  \def\citenamefont#1{#1}\fi
\expandafter\ifx\csname url\endcsname\relax
  \def\url#1{\texttt{#1}}\fi
\expandafter\ifx\csname urlprefix\endcsname\relax\def\urlprefix{URL }\fi
\providecommand{\bibinfo}[2]{#2}
\providecommand{\eprint}[2][]{\url{#2}}

\bibitem[{\citenamefont{{Regal} et~al.}(2004)\citenamefont{{Regal}, {Greiner},
  and {Jin}}}]{regal04}
\bibinfo{author}{\bibfnamefont{C.~A.} \bibnamefont{{Regal}}},
  \bibinfo{author}{\bibfnamefont{M.}~\bibnamefont{{Greiner}}},
  \bibnamefont{and} \bibinfo{author}{\bibfnamefont{D.~S.} \bibnamefont{{Jin}}},
  \bibinfo{journal}{Phys. Rev. Lett.} \textbf{\bibinfo{volume}{92}},
  \bibinfo{pages}{040403} (\bibinfo{year}{2004}).

\bibitem[{\citenamefont{{Zwierlein} et~al.}(2004)\citenamefont{{Zwierlein},
  {Stan}, {Schunck}, {Raupach}, {Kerman}, and {Ketterle}}}]{zwierlein04}
\bibinfo{author}{\bibfnamefont{M.~W.} \bibnamefont{{Zwierlein}}},
  \bibinfo{author}{\bibfnamefont{C.~A.} \bibnamefont{{Stan}}},
  \bibinfo{author}{\bibfnamefont{C.~H.} \bibnamefont{{Schunck}}},
  \bibinfo{author}{\bibfnamefont{S.~M.} \bibnamefont{{Raupach}}},
  \bibinfo{author}{\bibfnamefont{A.~J.} \bibnamefont{{Kerman}}},
  \bibnamefont{and}
  \bibinfo{author}{\bibfnamefont{W.}~\bibnamefont{{Ketterle}}},
  \bibinfo{journal}{Phys. Rev. Lett.} \textbf{\bibinfo{volume}{92}},
  \bibinfo{pages}{120403} (\bibinfo{year}{2004}).

\bibitem[{\citenamefont{{Bourdel} et~al.}(2004)\citenamefont{{Bourdel},
  {Khaykovich}, {Cubizolles}, {Zhang}, {Chevy}, {Teichmann}, {Tarruell},
  {Kokkelmans}, and {Salomon}}}]{bourdel04}
\bibinfo{author}{\bibfnamefont{T.}~\bibnamefont{{Bourdel}}},
  \bibinfo{author}{\bibfnamefont{L.}~\bibnamefont{{Khaykovich}}},
  \bibinfo{author}{\bibfnamefont{J.}~\bibnamefont{{Cubizolles}}},
  \bibinfo{author}{\bibfnamefont{J.}~\bibnamefont{{Zhang}}},
  \bibinfo{author}{\bibfnamefont{F.}~\bibnamefont{{Chevy}}},
  \bibinfo{author}{\bibfnamefont{M.}~\bibnamefont{{Teichmann}}},
  \bibinfo{author}{\bibfnamefont{L.}~\bibnamefont{{Tarruell}}},
  \bibinfo{author}{\bibfnamefont{S.~J.} \bibnamefont{{Kokkelmans}}},
  \bibnamefont{and}
  \bibinfo{author}{\bibfnamefont{C.}~\bibnamefont{{Salomon}}},
  \bibinfo{journal}{Phys. Rev. Lett.} \textbf{\bibinfo{volume}{93}},
  \bibinfo{pages}{050401} (\bibinfo{year}{2004}).

\bibitem[{\citenamefont{Paredes and Cirac}(2003)}]{paredes03}
\bibinfo{author}{\bibfnamefont{B.}~\bibnamefont{Paredes}} \bibnamefont{and}
  \bibinfo{author}{\bibfnamefont{J.~I.} \bibnamefont{Cirac}},
  \bibinfo{journal}{Phys. Rev. Lett.} \textbf{\bibinfo{volume}{90}},
  \bibinfo{pages}{150402} (\bibinfo{year}{2003}).

\bibitem[{\citenamefont{{Winkler} et~al.}(2006)\citenamefont{{Winkler},
  {Thalhammer}, {Lang}, {Grimm}, {Hecker Denschlag}, {Daley}, {Kantian},
  {B{\"u}chler}, and {Zoller}}}]{winkler06}
\bibinfo{author}{\bibfnamefont{K.}~\bibnamefont{{Winkler}}},
  \bibinfo{author}{\bibfnamefont{G.}~\bibnamefont{{Thalhammer}}},
  \bibinfo{author}{\bibfnamefont{F.}~\bibnamefont{{Lang}}},
  \bibinfo{author}{\bibfnamefont{R.}~\bibnamefont{{Grimm}}},
  \bibinfo{author}{\bibfnamefont{J.}~\bibnamefont{{Hecker Denschlag}}},
  \bibinfo{author}{\bibfnamefont{A.~J.} \bibnamefont{{Daley}}},
  \bibinfo{author}{\bibfnamefont{A.}~\bibnamefont{{Kantian}}},
  \bibinfo{author}{\bibfnamefont{H.~P.} \bibnamefont{{B{\"u}chler}}},
  \bibnamefont{and} \bibinfo{author}{\bibfnamefont{P.}~\bibnamefont{{Zoller}}},
  \bibinfo{journal}{Nature} \textbf{\bibinfo{volume}{441}},
  \bibinfo{pages}{853} (\bibinfo{year}{2006}).

\bibitem[{\citenamefont{Widera et~al.}(2004)\citenamefont{Widera, Mandel,
  Greiner, Kreim, Hansch, and Bloch}}]{widera04}
\bibinfo{author}{\bibfnamefont{A.}~\bibnamefont{Widera}},
  \bibinfo{author}{\bibfnamefont{O.}~\bibnamefont{Mandel}},
  \bibinfo{author}{\bibfnamefont{M.}~\bibnamefont{Greiner}},
  \bibinfo{author}{\bibfnamefont{S.}~\bibnamefont{Kreim}},
  \bibinfo{author}{\bibfnamefont{T.~W.} \bibnamefont{Hansch}},
  \bibnamefont{and} \bibinfo{author}{\bibfnamefont{I.}~\bibnamefont{Bloch}},
  \bibinfo{journal}{Physical Review Letters} \textbf{\bibinfo{volume}{92}},
  \bibinfo{eid}{160406} (\bibinfo{year}{2004}).

\bibitem[{\citenamefont{F\"olling et~al.}(2007)\citenamefont{F\"olling,
  Trotzky, Cheinet, Feld, Saers, Widera, M\"uller, and Bloch}}]{foelling07}
\bibinfo{author}{\bibfnamefont{S.}~\bibnamefont{F\"olling}},
  \bibinfo{author}{\bibfnamefont{S.}~\bibnamefont{Trotzky}},
  \bibinfo{author}{\bibfnamefont{P.}~\bibnamefont{Cheinet}},
  \bibinfo{author}{\bibfnamefont{M.}~\bibnamefont{Feld}},
  \bibinfo{author}{\bibfnamefont{R.}~\bibnamefont{Saers}},
  \bibinfo{author}{\bibfnamefont{A.}~\bibnamefont{Widera}},
  \bibinfo{author}{\bibfnamefont{T.}~\bibnamefont{M\"uller}}, \bibnamefont{and}
  \bibinfo{author}{\bibfnamefont{I.}~\bibnamefont{Bloch}},
  \bibinfo{journal}{Nature} \textbf{\bibinfo{volume}{448}},
  \bibinfo{pages}{1029} (\bibinfo{year}{2007}).

\bibitem[{\citenamefont{Anderlini et~al.}(2007)\citenamefont{Anderlini, Lee,
  Brown, Sebby-Strabley, Phillips, and Porto}}]{portopairs}
\bibinfo{author}{\bibfnamefont{M.}~\bibnamefont{Anderlini}},
  \bibinfo{author}{\bibfnamefont{P.~J.} \bibnamefont{Lee}},
  \bibinfo{author}{\bibfnamefont{B.~L.} \bibnamefont{Brown}},
  \bibinfo{author}{\bibfnamefont{J.}~\bibnamefont{Sebby-Strabley}},
  \bibinfo{author}{\bibfnamefont{W.~D.} \bibnamefont{Phillips}},
  \bibnamefont{and} \bibinfo{author}{\bibfnamefont{J.~V.} \bibnamefont{Porto}},
  \bibinfo{journal}{Nature} \textbf{\bibinfo{volume}{448}},
  \bibinfo{pages}{452} (\bibinfo{year}{2007}).

\bibitem[{\citenamefont{Rodriguez et~al.}(2007)\citenamefont{Rodriguez, Clark,
  and Jaksch}}]{rodriguez06}
\bibinfo{author}{\bibfnamefont{M.}~\bibnamefont{Rodriguez}},
  \bibinfo{author}{\bibfnamefont{S.~R.} \bibnamefont{Clark}}, \bibnamefont{and}
  \bibinfo{author}{\bibfnamefont{D.}~\bibnamefont{Jaksch}},
  \bibinfo{journal}{\pra} \textbf{\bibinfo{volume}{75}}, \bibinfo{eid}{011601}
  (\bibinfo{year}{2007}).

\bibitem[{\citenamefont{{Jaksch} et~al.}(1998)\citenamefont{{Jaksch}, {Bruder},
  {Cirac}, {Gardiner}, and {Zoller}}}]{jaksch98}
\bibinfo{author}{\bibfnamefont{D.}~\bibnamefont{{Jaksch}}},
  \bibinfo{author}{\bibfnamefont{C.}~\bibnamefont{{Bruder}}},
  \bibinfo{author}{\bibfnamefont{J.~I.} \bibnamefont{{Cirac}}},
  \bibinfo{author}{\bibfnamefont{C.~W.} \bibnamefont{{Gardiner}}},
  \bibnamefont{and} \bibinfo{author}{\bibfnamefont{P.}~\bibnamefont{{Zoller}}},
  \bibinfo{journal}{Phys. Rev. Lett.} \textbf{\bibinfo{volume}{81}},
  \bibinfo{pages}{3108} (\bibinfo{year}{1998}).

\bibitem[{\citenamefont{Kraus et~al.}()\citenamefont{Kraus, Wolf, Giedke, and
  Cirac}}]{kraus07}
\bibinfo{author}{\bibfnamefont{C.}~\bibnamefont{Kraus}},
  \bibinfo{author}{\bibfnamefont{M.}~\bibnamefont{Wolf}},
  \bibinfo{author}{\bibfnamefont{G.}~\bibnamefont{Giedke}}, \bibnamefont{and}
  \bibinfo{author}{\bibfnamefont{I.}~\bibnamefont{Cirac}},
  \bibinfo{howpublished}{in prep.}

\bibitem[{\citenamefont{Garc\'{i}a-Ripoll}(2006)}]{ripoll07}
\bibinfo{author}{\bibfnamefont{J.~J.} \bibnamefont{Garc\'{i}a-Ripoll}},
  \bibinfo{journal}{New J. Phys.} \textbf{\bibinfo{volume}{8}},
  \bibinfo{pages}{305} (\bibinfo{year}{2006}).

\bibitem[{\citenamefont{Verstraete et~al.}(2004)\citenamefont{Verstraete,
  Garcia-Ripoll, and Cirac}}]{verstraete04}
\bibinfo{author}{\bibfnamefont{F.}~\bibnamefont{Verstraete}},
  \bibinfo{author}{\bibfnamefont{J.~J.} \bibnamefont{Garcia-Ripoll}},
  \bibnamefont{and} \bibinfo{author}{\bibfnamefont{J.~I.} \bibnamefont{Cirac}},
  \bibinfo{journal}{Physical Review Letters} \textbf{\bibinfo{volume}{93}},
  \bibinfo{eid}{207204} (pages~\bibinfo{numpages}{4}) (\bibinfo{year}{2004}).

\bibitem[{\citenamefont{S.~Rapsch and Zwerger}(1999)}]{rapsch99}
\bibinfo{author}{\bibfnamefont{U.~S.} \bibnamefont{S.~Rapsch}}
  \bibnamefont{and} \bibinfo{author}{\bibfnamefont{W.}~\bibnamefont{Zwerger}},
  \bibinfo{journal}{Europhysics Letters (EPL)} \textbf{\bibinfo{volume}{46}},
  \bibinfo{pages}{559} (\bibinfo{year}{1999}).

\bibitem[{\citenamefont{{Altman} et~al.}(2004)\citenamefont{{Altman}, {Demler},
  and {Lukin}}}]{altman04}
\bibinfo{author}{\bibfnamefont{E.}~\bibnamefont{{Altman}}},
  \bibinfo{author}{\bibfnamefont{E.}~\bibnamefont{{Demler}}}, \bibnamefont{and}
  \bibinfo{author}{\bibfnamefont{M.~D.} \bibnamefont{{Lukin}}},
  \bibinfo{journal}{\pra} \textbf{\bibinfo{volume}{70}},
  \bibinfo{pages}{013603} (\bibinfo{year}{2004}).

\bibitem[{\citenamefont{{F{\"o}lling} et~al.}(2005)\citenamefont{{F{\"o}lling},
  {Gerbier}, {Widera}, {Mandel}, {Gericke}, and {Bloch}}}]{foelling05}
\bibinfo{author}{\bibfnamefont{S.}~\bibnamefont{{F{\"o}lling}}},
  \bibinfo{author}{\bibfnamefont{F.}~\bibnamefont{{Gerbier}}},
  \bibinfo{author}{\bibfnamefont{A.}~\bibnamefont{{Widera}}},
  \bibinfo{author}{\bibfnamefont{O.}~\bibnamefont{{Mandel}}},
  \bibinfo{author}{\bibfnamefont{T.}~\bibnamefont{{Gericke}}},
  \bibnamefont{and} \bibinfo{author}{\bibfnamefont{I.}~\bibnamefont{{Bloch}}},
  \bibinfo{journal}{Nature} \textbf{\bibinfo{volume}{434}},
  \bibinfo{pages}{481} (\bibinfo{year}{2005}).

\bibitem[{\citenamefont{Eckert et~al.}()\citenamefont{Eckert, Romero-Isart,
  Rodr{\'\i}guez, Lewenstein, Polzik, and Sanpera}}]{eckert07}
\bibinfo{author}{\bibfnamefont{K.}~\bibnamefont{Eckert}},
  \bibinfo{author}{\bibfnamefont{O.}~\bibnamefont{Romero-Isart}},
  \bibinfo{author}{\bibfnamefont{M.}~\bibnamefont{Rodr{\'\i}guez}},
  \bibinfo{author}{\bibfnamefont{M.}~\bibnamefont{Lewenstein}},
  \bibinfo{author}{\bibfnamefont{E.}~\bibnamefont{Polzik}}, \bibnamefont{and}
  \bibinfo{author}{\bibfnamefont{A.}~\bibnamefont{Sanpera}},
  \bibinfo{note}{arXiv:0709.0527}.

\end{thebibliography}
\end{document}